\documentclass[12pt]{article}

\usepackage{amsmath,amsfonts,eurosym,ulem,graphicx,caption,color,setspace,sectsty,comment,footmisc,caption,pdflscape,subfigure,array}
\usepackage[T1]{fontenc}
\usepackage{newtxtext,newtxmath}
\usepackage{microtype}
\usepackage[colorlinks=true, linkcolor=blue, urlcolor=blue, citecolor=blue]{hyperref}
\usepackage[round]{natbib}
\usepackage{longtable}
\usepackage{tabularx}
\usepackage{booktabs}
\renewcommand{\arraystretch}{1.2}  % Optional: improves readability
\usepackage{float}
\usepackage[most]{tcolorbox}

\usepackage{changepage}
\usepackage[margin=2.5cm]{geometry}
\usepackage{adjustbox}
\usepackage{float} % if you use [H] (not required here)
\usepackage{epigraph}

\newcolumntype{L}[1]{>{\raggedright\let\newline\\arraybackslash\hspace{0pt}}m{#1}}
\newcolumntype{C}[1]{>{\centering\let\newline\\arraybackslash\hspace{0pt}}m{#1}}
\newcolumntype{R}[1]{>{\raggedleft\let\newline\\arraybackslash\hspace{0pt}}m{#1}}

\begin{document}

\title{Strategic Integration of Artificial Intelligence in the C-Suite: The Role of the Chief AI Officer}
\author{Marc Schmitt\\[-0.6em] \normalsize University of Oxford}
\date{}
\maketitle

\begin{abstract}

\noindent The integration of Artificial Intelligence (AI) into corporate strategy has become critical for organizations seeking to maintain competitive advantage in the digital age. Although organizations increasingly rely on AI as a strategic and organizational resource, existing C-suite roles remain only partially equipped to govern, integrate, and leverage it coherently at the enterprise level. Organizations vary in their responses. Some create a dedicated Chief AI Officer (CAIO), others extend existing mandates into hybrid roles, and still others coordinate AI through federated structures. This paper develops a role-design theory to explain this variation. I identify three properties that distinguish AI systems from earlier cross-cutting enterprise technologies—distributed accountability for judgment, upstream governance, and non-stationarity—and three configurations through which organizations respond: concentrated extension, distributed extension, and role creation. Four propositions specify when a dedicated CAIO emerges, what form an organization's response takes, when the role is effective, and how configurations evolve. Drawing on the first archival census of executive AI leadership, I find that the role has diffused sharply since 2023, that role creation and mandate extension are advancing in near parity, that adoption concentrates where digital leadership structures already exist, and that executive tenures contradict practitioner accounts of a transient role. The paper contributes to research on executive leadership, AI governance, and the design of AI-driven organizations.
\\
\vspace{0in}
\vspace{0in}

\bigskip
\end{abstract}
\setcounter{page}{0}
\thispagestyle{empty}
\pagebreak \newpage

\onehalfspacing

\section{Introduction} \label{sec:intro}

Artificial intelligence (AI) has become a general-purpose force reshaping organizational decision making, innovation, and competition \citep{Goldfarb2023CouldPostings}. Recent advances in multimodal, generative, and agentic AI have expanded its role across the full spectrum of organizational capabilities, from products and processes to coordination and strategic choice \citep{Hillebrand2025ManagingFramework}. As a result, AI is increasingly embedded in the organizational architecture through which firms create value, organize work, and compete \citep{Kellogg2020AlgorithmsControl,Anthony2023CollaboratingWork,Kemp2024CompetitiveAI,Matthews2025ReviewAI}. In doing so, it redistributes agency and accountability across human and algorithmic actors, relocating a share of organizational judgment into computational systems \citep{Murray2021HumansOrganizations}.

This shift creates a governance problem at the executive level. AI cuts across strategy, operations, innovation, technology, and risk, yet organizations often lack a clear structure for enterprise-wide AI leadership. Responsibility is dispersed across multiple senior roles. Although the Chief Information Officer (CIO), Chief Technology Officer (CTO), Chief Digital Officer (CDO), and Chief Strategy Officer (CSO) each address parts of the AI challenge, none is designed to integrate it across the full range of governance, transformation, capability-building, and strategic coordination demands it creates \citep{Menz2014ChiefTeams, Bendig2022WhenPresence, Firk2021ChiefRole}. These demands arise because AI combines cross-functional reach, strategic consequences, and infrastructural dependence in ways that strain existing role boundaries \citep{Kemp2024CompetitiveAI, Krakowski2023ArtificialAdvantage}. The result is an executive governance gap: a misalignment between the enterprise-wide consequences of AI and the distribution of authority, accountability, and expertise across existing C-suite roles \citep{Raisch2021ArtificialParadox}. Organizations are beginning to respond to this gap, but their responses vary. Some create dedicated Chief AI Officer positions. Others extend existing mandates such as the CDO into hybrid Chief Data and AI Officer or Chief Digital and AI Officer roles. Others coordinate AI through committees and federated governance structures.

\begin{figure}[htbp]
\centering
\caption{The Rise of Executive AI Leadership, 2018--2026}
\label{fig:rise}
\includegraphics[width=0.85\textwidth]{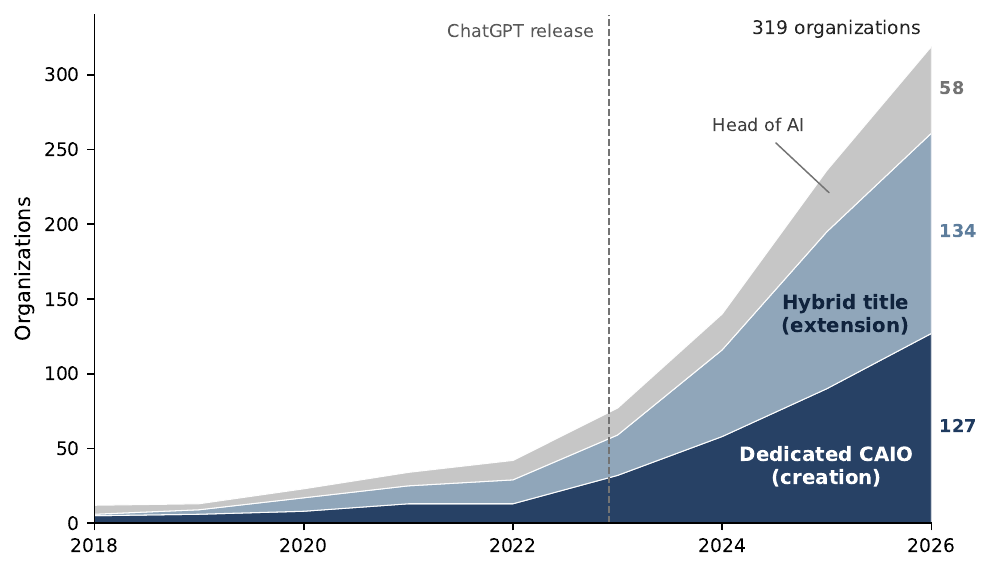}
\par\vspace{4pt}
{\footnotesize\raggedright\setlength{\parindent}{0pt}%
\textit{Note.} Global census of executive AI leadership roles constructed from BoardEx organization composition files (North America, Europe, UK, Rest of World). Bands show organizations with an active role by configuration: dedicated Chief AI Officer (role creation), hybrid AI titles such as Chief Data \& AI Officer (concentrated extension), and senior sub-executive Head of AI positions. Dashed line marks the release of ChatGPT (November 2022). Right-margin values give the August 2026 count of organizations holding each configuration; an organization holding more than one configuration at a date is counted in each band, so band totals (319 in August 2026) exceed unique organizations (310). The 2026 value is through August 1. Section~\ref{sec:evidence} details sources and construction.\par}
\end{figure}

Enterprise technologies have crossed functional boundaries before. Enterprise resource planning reshaped operations across every function; the internet, cloud computing, and mobile platforms each transformed how firms compete. Yet none of them produced a lasting new seat in the C-suite: organizations absorbed each wave into the mandates of existing executives. AI is breaking that pattern. Figure~\ref{fig:rise} documents the break: from single digits before 2020 to 258 organizations worldwide with a chief-level AI executive by August 2026 (310 including senior Head of AI positions), split almost evenly between newly created Chief AI Officer positions and extended hybrid mandates. Importance alone cannot explain why AI, unlike its predecessors, produced new executive roles, since earlier technologies were important too. What differs is the nature of the systems. AI systems perform tasks that involve organizational judgment. Their outputs depend on what they have learned, and their behavior can change after deployment. Distributed accountability for judgment, upstream governance, and non-stationarity are not properties of markets, executives, or institutional fields; they are properties of the technology itself. Explaining this resistance to absorption requires a theory that connects the properties of AI systems to the organization of executive authority. This paper develops that theory.

This paper asks why organizations facing similar AI pressures adopt such different executive responses, and under what conditions a dedicated Chief AI Officer (CAIO) emerges as the appropriate design. I argue that this variation is systematic. It follows from three properties that distinguish AI from earlier cross-cutting enterprise technologies and from the structure of the C-suite each organization inherits. I develop a role-design theory of the CAIO that traces these properties to the executive design problems they create, explains three configurations through which organizations respond, and specifies the conditions under which each configuration is most likely to emerge, persist, or transition.

The paper makes three main contributions. First, I develop a theory of executive technology leadership that starts from the properties of the technology being governed. Building on research that traces organizing consequences to the properties of digital artifacts \citep{Kallinikos2013AmbivalentOntology}, I identify three properties that distinguish AI systems from earlier enterprise technology---distributed accountability for judgment, upstream governance, and non-stationarity---and show how their joint presence generates an executive design problem that existing roles cannot fully absorb. Theories of executive attention, fashion, and power explain why firms respond to important technologies. This paper extends the explanation by asking how the properties of AI shape whether firms create new executive roles or expand existing ones. Second, I theorize three configurations through which organizations respond to this problem: concentrated extension, distributed extension, and role creation. The first two preserve the existing C-suite architecture. The third establishes a dedicated CAIO. Third, I develop four propositions that specify the conditions under which each configuration is most likely to emerge, the factors that shape which configuration a given firm adopts, the conditions under which a dedicated CAIO is effective once created, and the trajectories along which configurations evolve as AI's organizational footprint changes.

\section{Why AI Demands Distinct Leadership} \label{sec:AIdemands}

Following \citet{Alvesson2011GeneratingProblematization}, this paper problematizes the assumption that existing executive roles are adequate for enterprise-wide AI leadership. It argues that AI creates a distinct leadership problem that requires rethinking how firms structure their top management teams.

\citet{Bendig2022WhenPresence} show that the presence of CIOs in the top management team is shaped by environmental, structural, and strategic pressures. These pressures can be summarized as follows: environmental turbulence increases the need for executive-level technology oversight; structural gaps in digital expertise create demand for new leadership roles; and strategy increasingly depends on the integration of technological and business priorities. I build on this logic to argue that similar but stronger pressures now support the emergence of a Chief AI Officer, especially in light of the economic, organizational, and competitive shifts AI is producing.

Research suggests that the mere presence of AI-related technologies and technical expertise does not ensure that firms will successfully integrate AI or translate it into meaningful innovation. Consistent with the argument developed by \citet{Chen2021MakingPerspective}, successful AI integration and innovation may require executive leadership that can mobilize organizational attention to AI, coordinate initiatives across business and technical domains, align AI efforts with strategic priorities, and scale fragmented experimentation at the enterprise level.

Adjacent executive-role research shows that specialized functional leaders can shape organization-wide outcomes when they enter the top management team \citep{Menz2012FunctionalAgenda, Zhu2025TopExecutivesInnovation}. For example, \citet{Gao2025ChiefInnovation} show that the presence of a Chief Information Security Officer (CISO) on the top management team (TMT) can enhance innovation through strategic risk reduction, technology enablement, and the protection of innovation-related assets. A similar dynamic appears for non-technology roles. As \citet{Fu2020ChiefIrresponsibility} show, the presence of a chief sustainability officer on the top management team channels executive attention to a previously dispersed domain, consistent with the upper-echelons perspective and the attention-based view. This reinforces the broader point that specialized functional executive roles can matter at the level of firm strategy and value creation.

Research on technology executives has chronicled this structural evolution one role at a time: the antecedents and strategic positioning of the CIO \citep{Banker2011CIOPerformance1, Bendig2022WhenPresence, Chen2021MakingPerspective}, the emergence of the CDO \citep{Firk2021ChiefRole, Kunisch2022ChiefDeterminants}, the organizational effects of the CISO \citep{Gao2025ChiefInnovation}, and most recently the digital orientation of CEOs and boards \citep{Filatotchev2025DigitalOrientation}. What this stream has lacked is a general account of when a technology domain gives rise to a dedicated executive role and when it is instead absorbed into existing mandates. The framework developed in this paper supplies that account. Its absorption-versus-differentiation logic connects earlier roles to the problems they addressed---enterprise computing for the CIO, digital business models for the CDO, and security exposure for the CISO---and specifies when AI warrants a durable executive role. The CAIO is thus the focal case of a broader theory of executive technology leadership that the literature on individual technology roles has been building toward.

\subsection{AI Is Everywhere, Yet No One Is Responsible}

Artificial intelligence is now embedded in core organizational activities \citep{Hillebrand2025ManagingFramework, Kemp2024CompetitiveAI, Brynjolfsson2025GenerativeWork}. As a result, AI is shifting from a functional tool to a foundational organizational asset, transforming decision environments across the enterprise and placing pressure on existing management and control systems \citep{Boyac2024HumanLimitations, Shrestha2019OrganizationalIntelligence}. Yet despite AI's growing ubiquity and strategic centrality, responsibility for AI leadership often remains fragmented and ambiguous. This creates a fundamental paradox: AI is increasingly enterprise-wide, but no single executive is formally accountable for orchestrating it across strategic, organizational, technical, and governance dimensions. The result is an executive governance gap that can produce inconsistent deployment, duplicated efforts, and heightened exposure to regulatory, ethical, and reputational risks \citep{Raisch2021ArtificialParadox}.

This fragmentation is especially problematic because AI differs from many earlier generations of enterprise technology. AI systems are dynamic, probabilistic, and often non-deterministic; they learn from data, operate with limited transparency, and may generate emergent effects that are difficult to anticipate ex ante \citep{deVericourt2023IsKnow}. These characteristics create novel forms of uncertainty and accountability that cut across functional boundaries and are not easily contained within existing leadership structures \citep{Raisch2021ArtificialParadox, Waardenburg2022InAlgorithms}. External pressures further intensify this challenge. Regulatory developments, public-sector mandates, and growing stakeholder expectations signal the need for clearer accountability in how organizations develop and deploy AI.

\subsection{Problematizing Existing C-Suite Roles}

Although executive leadership structures have evolved in response to technological change, these roles often face limits when confronted with the enterprise-wide strategic, organizational, and governance demands created by AI. To clarify these limits, I problematize existing C-suite roles across three dimensions: environmental, structural, and strategic. Each corresponds to one level at which AI reshapes organizing: the AI economy, the AI organization, and competition in the age of AI.

\textbf{Environmental limitations.} The AI economy intensifies external pressures on firms to respond to technological breakthroughs, regulatory developments, and shifting stakeholder expectations with speed and coherence. As AI becomes a source of innovation, efficiency, and competitive adaptation, firms require leadership capable of interpreting these pressures and translating them into enterprise-wide responses. Competitive dynamics theory suggests that such environments reward rapid and coordinated adaptation \citep{Chen1994CompetitiveFramework}. Yet existing roles such as the CIO and CTO are typically bounded by functional mandates centered on infrastructure, systems, or product domains. They are not institutionally designed to align decentralized AI experimentation, cross-business-unit innovation, and enterprise-wide strategic adaptation. This limitation becomes more acute as AI development diffuses across the organization through accessible tools and local initiatives, increasing the risks of duplication, fragmentation, and governance inconsistency. The result is an environmental leadership gap: firms face AI-driven external turbulence, while existing roles are configured primarily for narrower functional or technical mandates.

\textbf{Structural limitations.} The AI organization creates structural demands that exceed conventional technology leadership. AI systems raise issues of bias, transparency, accountability, and compliance that cut across organizational domains. While existing executives may address selected parts of this landscape--the CIO may oversee data and systems, the CTO technology development, the CDO digital transformation, and the CSO strategic direction--no single role is consistently designed to coordinate AI across these interdependent arenas. The proliferation of adjacent roles such as Chief Data Officer, Chief Transformation Officer, and Chief Digital Officer further reveals the fragmentation of responsibilities once concentrated under the CIO. Although this differentiation increases specialization, it also creates structural blind spots for AI, which simultaneously implicates governance, innovation, and organizational redesign.

Among adjacent roles, the Chief Digital Officer comes closest to addressing this challenge, though prior research suggests that the role has typically been oriented toward digital transformation rather than enterprise-wide AI governance \citep{Wessel2021Unpacking}. \citet{Firk2021ChiefRole} theorize the CDO as a centralized digital transformation role responsible for a holistic digital strategy and the associated change effort, and identify its core benefits in accelerating and coordinating digital transformation. \citet{Kunisch2022ChiefDeterminants} further characterize the CDO as a distinct executive role with a company-wide digital mandate centered on moving the organization toward a digital mindset and coordinating digital initiatives across the firm. This makes the CDO an important integrator of digital transformation, and one whose mandate can in principle be extended to encompass AI, as the emergence of hybrid Chief Data and AI Officer and Chief Digital and AI Officer titles suggests. Whether such extension is sufficient depends on conditions developed in Section~\ref{sec:framework}: AI differs from earlier digital technologies in being dynamic, probabilistic, and often non-deterministic, creating cross-domain issues of governance, accountability, and decision rights that may exceed the boundaries within which the CDO mandate was originally designed to operate. The result is a structural coordination gap whose severity varies with the extent to which AI's distinctive properties intensify. In practice, incumbent-role configurations vary more widely than the CDO-focused literature suggests, with AI mandates also extending into Chief Information Officer, Chief Technology Officer, and Chief Data Officer roles, among others.

\textbf{Strategic limitations.} AI also challenges the strategic logic of the modern C-suite. In the age of AI, data assets, algorithmic capabilities, and digital infrastructures increasingly shape competitive advantage and innovation trajectories. Thus, AI is becoming constitutive of strategy itself \citep{Kemp2024CompetitiveAI, Bichler2026AgenticMarkets}. Yet most executive roles remain rooted in a legacy separation between ``business'' and ``technology'' \citep{Henderson1993StrategicAlignment}. The Chief Strategy Officer may define direction but typically lacks the technical fluency needed to evaluate AI affordances, constraints, and path dependencies. The CTO may understand technological possibilities but usually lacks the mandate to align them across business units, governance structures, and long-range organizational priorities. Even where CIOs participate in strategic decision making, their authority is often still anchored in infrastructure and enterprise systems \citep{Banker2011CIOPerformance1, Bendig2023AttentionTeam}. The result is a strategic integration gap: firms require leadership capable of integrating deep technical understanding with cross-functional authority and long-term strategic vision, and no established C-suite role is consistently designed to perform that function for AI without significant adaptation.

\subsection{Role Extension and Role Creation}
\label{subsec:two-responses}

Faced with the executive demands AI generates, organizations have two broad design responses available. The first is \textit{role extension}, absorbing AI into the mandate of one or more existing executive roles. The second is \textit{role creation}, establishing a dedicated AI executive position, typically titled Chief AI Officer. Any theory of the CAIO must engage seriously with role extension as a competing organizational solution. It is the historically dominant response to cross-cutting technologies, which organizations absorbed into the mandates of the CIO, CTO, and CDO instead of creating new roles.

Role extension takes two forms. \textit{Concentrated extension} assigns AI to a single incumbent executive, most often a data or digital executive, producing hybrid titles such as Chief Data and AI Officer or Chief Digital and AI Officer. \textit{Distributed extension} assigns AI across multiple existing roles, coordinated through committees, councils, or federated governance structures in which the CIO, CTO, CDO, Chief Risk Officer, and Chief Legal Officer share AI-related decisions according to domain, a structure with a board-level analogue in technology committees \citep{Clark2026TechnologyCommittees}. Both forms are extensions in the same theoretical sense. They preserve the existing C-suite architecture and absorb AI into mandates designed for other problems. Distributed extension differs from concentrated extension in the breadth of absorption. Its underlying logic remains the same.

Role extension has three structural advantages. It conserves executive headcount, which matters in organizations where the size of the C-suite is itself a governance concern. It builds on established authority, reporting lines, and institutional credibility, allowing the incumbent executive to act more quickly on AI. And it avoids the jurisdictional disputes that follow the creation of any new senior role, particularly one whose remit overlaps with existing positions. For organizations in which AI is consequential but not yet enterprise-defining, role extension is the more efficient design.

Role creation, the establishment of a dedicated CAIO, is the response this paper theorizes. Role extension becomes inadequate under specific conditions, and those conditions become more common as AI becomes more deeply embedded in organizational activity. The reason is structural. Role extension preserves the architecture of the role or roles it extends, adding AI to portfolios whose authority, accountability, and capability sets were designed for different problems. When AI is treated as an additional infrastructural or transformational concern within the CDO mandate, or as a distributed concern coordinated across legal, risk, and technology functions, it is governed using the integration and coordination tools those mandates already possess. This holds as long as AI's consequences remain bounded, with deployment limited to discrete applications, narrow accountability exposure, and contained governance demand. As AI moves toward enterprise-wide deployment, distributed accountability for substantive judgments, and continuous non-stationary governance, the gap widens between the problem and the tools the extended mandate possesses. These are the conditions under which the jurisdictional ambiguity, accountability externalities, and orchestration demands developed in Section~\ref{subsec:design-problems} exceed what an extended mandate can resolve.

Three failure modes follow. The first is \textit{governance lag}: extended attention is divided between the legacy mandate and the new AI mandate, producing slower response to AI-specific risks than a dedicated role would generate. Under distributed extension, lag is compounded by coordination overhead across roles with different incentives and time horizons. The second is \textit{capability dilution}: the capability bundles required for legacy mandates, such as digital transformation execution, infrastructure management, and risk compliance, differ from those required for the AI mandate, such as judgment governance, model accountability, and ongoing oversight of changing AI systems. Consolidating both in an existing role tends to under-resource one or the other. The third is \textit{legitimacy ambiguity}: stakeholders inside and outside the organization remain uncertain about which executive is answerable for AI-related decisions, which complicates accountability in the domain where it is most contested.

\subsection{The Executive Governance Gap}
\label{subsec:gov-gap}

Role extension absorbs AI into mandates designed for other problems, and it does so successfully only while AI's consequences remain bounded. Once AI's distinctive properties intensify, extension encounters the failure modes just described, and no existing role is configured to resolve them. The result is an executive governance gap: existing arrangements do not adequately integrate technical understanding, organizational coordination, and strategic direction for AI at the enterprise level. This gap is visible across the three levels introduced in Section~\ref{sec:AIdemands}: the AI economy, the AI organization, and competition in the age of AI.

Existing roles each absorb important fragments of the AI challenge, but none is consistently designed to orchestrate AI across all three dimensions. The environmental, structural, and strategic dimensions locate where AI generates leadership demands, while the mechanism developed in the next section explains why existing roles cannot absorb them. What is needed is an executive arrangement with the mandate to translate AI's distinctive properties into coherent organizational action. Whether that arrangement is a dedicated role or an extended mandate depends on conditions the next section specifies. Section~\ref{sec:framework} develops the CAIO Framework by explaining how the gap arises, identifying the executive design problems it creates, and specifying when a dedicated role becomes warranted.

\section{The CAIO Framework: A Theory of Executive AI Leadership} \label{sec:framework}

The CAIO Framework explains how AI's distinctive properties create executive design problems and shape firms' choices among three leadership configurations. It also identifies the functions each configuration must perform, with the dedicated CAIO as the focal case. Figure~\ref{fig:CAIO} summarizes these relationships. The section begins by examining AI as an organizational object.

\subsection{AI as an Organizational Object}

To translate the preceding argument into a concrete executive role design, it is necessary to clarify how AI should be understood from an organizational standpoint. For the purposes of this paper, AI is defined to include computational systems that support or perform activities such as prediction, classification, recommendation, generation, detection, and decision making. These systems range from rule-based logic and conventional machine learning to generative AI, including large language models, and increasingly autonomous agentic systems \citep{LeCun2015DeepLearning}. From an organizational perspective, however, AI is not a monolithic technology. Rather, it is better understood as a layered organizational object that spans applications, infrastructure, and governance mechanisms. At the application layer, AI appears in systems that support prediction, generation, recommendation, detection, and decision support \citep{Raisch2021ArtificialParadox}. At the infrastructure layer, it depends on data pipelines, model-development environments, deployment architectures, and monitoring capabilities \citep{Hillebrand2025ManagingFramework}. At the governance layer, it requires mechanisms for explainability, risk control, fairness, privacy, compliance, and accountability \citep{Chen2023AlgorithmsAttributions, Tarafdar2023AlgorithmsWork}. Starting from the technology itself follows a research tradition in which the properties of digital artifacts are theorized as a source of their organizational consequences \citep{Kallinikos2013AmbivalentOntology}.

This layered character helps explain why AI leadership cannot be confined to a single functional domain. Because AI's applications, infrastructure, data, and oversight are distributed across the firm, they create recurring coordination and accountability problems that do not align neatly with any single executive boundary. This view is reinforced by recent work suggesting that AI's organizational significance extends beyond its functional applications to include changes in information processing, delegation dynamics, and the coordination of human and artificial agents \citep{Bauer2023ExplAIned:Processing, vonZahn2025KnowingMetacognition, Grisold2025GuardrailsPredictability}.

\subsection{The Mechanism: Three Properties of AI}
\label{subsec:mechanism}

Cross-cutting enterprise technologies are not new. Enterprise resource planning systems, internet infrastructure, cloud computing, and mobile platforms each crossed functional boundaries, reshaped operating models, and demanded executive attention. None produced a durable, distinct C-suite role. Organizations absorbed them into the mandates of the CIO, the CTO, and later the CDO. A theory of the CAIO must therefore identify what is different about AI as an organizational object. If AI resembles these earlier technologies in its essential organizational properties, then role extension should remain the efficient response, and a dedicated CAIO should not emerge as a stable position.

The argument I develop here rests on a single distinction. Earlier enterprise technologies were instruments of execution. They stored, transmitted, processed, or automated information according to specifications that humans set in advance. Accountability for the content of a decision stayed with the human actors who used these instruments, even when the technology mediated their action. The executive design problem these technologies created was one of infrastructural integration: aligning systems, data, and processes across functional boundaries. The CIO and the CDO emerged as integrative roles because the underlying problem was integrative. Those earlier systems, however editable, distributed, and generative, executed logic that humans had specified, so the demands they placed on organizations remained demands of integration.

AI changes the content of the problem. Contemporary AI systems generate predictions, recommendations, classifications, and increasingly autonomous actions whose substantive content no human specifies in advance. Their outputs follow from learned representations whose internal logic is often opaque even to the people who build them \citep{Lebovitz2022ToDiagnosis, Waardenburg2022InAlgorithms, deVericourt2023IsKnow}. A portion of organizational judgment therefore moves from human actors into computational artifacts \citep{Murray2021HumansOrganizations}. This relocation gives rise to three properties that together distinguish AI from earlier cross-cutting technologies.

The first property is \textit{distributed accountability for judgment}. When an AI system contributes to a consequential decision, such as granting credit, ranking a job candidate, or recommending a course of treatment, the question of who answers for that decision cannot be settled by asking who operated the system. Responsibility spreads across the developers of the model, the curators of its training data, the designers of its deployment context, and the human actors who accept or override its output. Earlier technologies raised accountability questions about systems. AI raises accountability questions about judgments, and existing executive mandates do not clearly allocate responsibility for those judgments. The difficulty deepens when the human actors positioned to check AI outputs are themselves miscalibrated, since reliance on confident-seeming systems can breed a misplaced certainty that erodes their independent judgment \citep{Allen2022Algorithm-AugmentedAversion, Leonardi2026KnowingPlanning}.

The second property is \textit{upstream governance}. For earlier enterprise technologies, governance attached to deployment: who holds access, how data moves, and how service levels are maintained. For AI, governance must also address how the system is trained and configured. What data trained the model, what objective it was optimized against, what biases it encodes, and under what conditions its performance degrades become governance questions in their own right. These are constitutive choices that shape what the organization will later do and answer for. They cannot be delegated cleanly to a technical function, because they bear directly on strategy. For AI, the division of labor between strategy and technology that served earlier technologies breaks down.

The third property is \textit{non-stationarity}. Enterprise systems, once deployed, behave as specified. AI systems drift. Their performance changes as the world changes, as the distribution of inputs shifts, and as users adapt their behavior to the system's presence. Governance through one-time architectural decisions, of the kind the CIO and CDO mandates were built around, does not hold a moving target in place. AI requires continuous executive attention, which favors a role whose primary charge is AI rather than a role for which AI is one portfolio among several. The challenge is compounded by the difficulty of specifying coordination requirements in advance, which has led recent work to emphasize guardrails and design for predictability in settings where human and AI activity cannot be fully anticipated ex ante \citep{Grisold2025GuardrailsPredictability}.

These three properties define a \textit{judgment-governance problem}. AI still creates integration problems, as earlier cross-cutting technologies did. What distinguishes it is the judgment-governance problem layered on top, which existing roles are not configured to resolve. Earlier cross-cutting technologies posed integration problems that existing roles could absorb. The judgment-governance problem is harder to absorb, and it grows harder as each property intensifies. When AI distributes accountability for substantive judgments, moves governance upstream to how models are built, and keeps changing after deployment, the executive arrangements that worked for earlier technologies face a widening gap between the problem they were designed for and the problem they now confront. The next subsection traces how these three properties manifest as concrete design problems at the level of the C-suite.

\subsection{Executive Design Problems}
\label{subsec:design-problems}

The three properties developed above are properties of AI as an organizational object. They do not become executive concerns until they manifest as problems that the C-suite must resolve. In this subsection, I develop three such problems: jurisdictional ambiguity, accountability externalities, and attention and orchestration. The mapping between properties and problems is not one-to-one. Each property contributes to more than one problem, and each problem draws on more than one property. What matters for the theory is that the three properties acting together generate all three problems at once, and that no existing executive role is configured to resolve the three jointly.

\textbf{Jurisdictional ambiguity.} AI spans strategy, technology, data, operations, and risk. Because it relocates judgment and raises governance to the level of model constitution, decisions about AI implicate several executive domains at the same time. A decision to deploy a model touches technical feasibility, which belongs to the CTO; data stewardship, which belongs to the CIO or the CDO; strategic positioning, which belongs to the CSO; and regulatory exposure, which belongs to legal and risk. Each executive holds partial authority over the decision, and none holds authority over the whole of it. Earlier technologies produced overlaps of this kind, but the overlaps were resolvable because authority could be partitioned by system or by process \citep{Joseph2020OrganizationalResearch}. AI resists clean partition, because the judgment a model performs cannot be assigned to a single functional owner. The result is a region of overlapping but incomplete authority in which decisions stall, default to inaction, or get made by whichever executive moves first.

\textbf{Accountability externalities.} AI generates exposures that fall on the organization as a whole rather than on the function that deploys the system. A biased model, an opaque decision, a compliance failure, or a reputational incident can impose costs that extend well beyond the unit that built or ran the system. Because accountability for judgment is distributed and governance sits upstream, no single function fully internalizes the cost of an AI failure. Local units capture the benefits of their own AI initiatives while the costs of failure are borne collectively. Left unaddressed, this produces systematic underinvestment in governance relative to deployment, because the executives best positioned to move quickly on AI are often not the executives who bear the consequences when it goes wrong.

\textbf{Attention and orchestration.} AI does not resolve into a fixed set of decisions that an executive can settle and move past. Because it is non-stationary, AI requires sustained prioritization across competing use cases, shared capabilities, infrastructure, external partners, and business units with divergent interests. Models must be monitored, retired, and replaced. Investments must be sequenced. Conflicts between units pursuing incompatible AI agendas must be arbitrated on a recurring basis. This is an orchestration problem, and orchestration consumes executive attention. Existing executives can supply attention to AI only by withdrawing it from the mandates they already hold. As AI's footprint grows, the attention an adequate response requires begins to exceed what any incumbent can spare without neglecting the role they were appointed to perform.

These three problems are the executive-level expression of AI's distinctive properties. They are not implementation difficulties that better project management would dissolve. They arise from the structure of AI as an organizational object, and they persist regardless of how competently any individual AI initiative is run. An organization confronting all three at once faces a genuine design choice about how to allocate executive authority over AI \citep{Joseph2025OrganizationDesign}. The next subsection develops the configurations available to it and identifies the conditions under which each becomes the organization's response.

\subsection{Configurations and Functions}
\label{subsec:framework}

\begin{figure}[htbp]
\centering
\caption{The CAIO Framework}
\label{fig:CAIO}
\includegraphics[width=\textwidth]{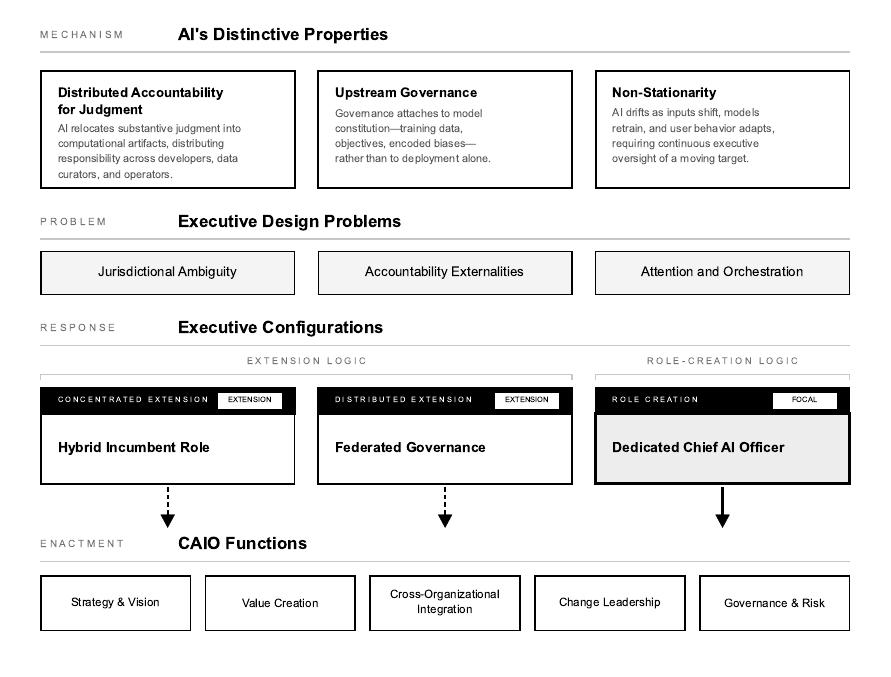}
\par\vspace{4pt}
{\footnotesize\raggedright\setlength{\parindent}{0pt}%
\textit{Note.} The framework reads top-down across four layers. The dedicated
CAIO (shaded) is the focal configuration. The five functions are
configuration-invariant demands: dashed arrows indicate that extension
configurations must perform the same functions within shared or dispersed incumbent mandates,
while the solid arrow marks their concentration in the dedicated CAIO mandate.\par}
\end{figure}

Figure~\ref{fig:CAIO} connects AI's distinctive properties to executive design problems, alternative leadership configurations, and the functions each configuration must perform. Concentrated and distributed extension adapt existing mandates, whereas role creation establishes a dedicated CAIO. The framework treats the dedicated role as the focal configuration because it is the one designed specifically for the judgment-governance problem.

The functions the dedicated role must perform follow from the three design problems. No single function maps neatly onto a single problem. Each problem requires particular functions, and only the full set can address the three together. Jurisdictional ambiguity, in which authority over AI is divided among incumbents who each hold partial claims, requires a function that coordinates AI across departments, functions, infrastructures, and partners; this is the integration function. Accountability externalities, in which the costs of AI failure fall on the organization while the benefits accrue locally, require a function that internalizes those costs through enterprise-level oversight of risk, ethics, compliance, privacy, and security; this is the governance and risk function. Attention and orchestration, in which AI demands sustained prioritization across competing uses under non-stationary conditions, calls for two functions: one that sets enterprise-level direction and maintains coherence between strategy and AI, and one that identifies, prioritizes, and scales the AI investments through which value is realized; these are the strategy and vision function and the value creation function. Finally, because AI relocates judgment into systems that people must learn to interpret, calibrate, and delegate to \citep{Choudhary2025HumanAIEnsembles}, resolving these problems in practice requires a function that prepares the organization for human--AI collaboration; this is the change leadership function. Thus, the five functions follow from the joint requirements of the three design problems. These functions arise from the design problems, so all three configurations must perform them. What differs is how responsibility for them is allocated. Role creation concentrates all five in a single mandate, while concentrated extension bundles them into a mandate shared with a legacy portfolio and distributed extension disperses them across several incumbents---the structural root of the capability dilution identified in Section~\ref{subsec:two-responses}. The mandate-content evidence in Section~\ref{subsec:functions} is consistent with this invariance: announcements of hybrid appointments reference the same five functions at rates similar to dedicated appointments.

Taken together, the four layers position the dedicated CAIO as an integrative executive role that translates AI's distinctive properties into strategic direction, organizational transformation, and institutional accountability. The functions in the fourth layer describe what that role does once an organization has chosen to create it. They do not imply that every organization should create a dedicated CAIO. The propositions developed in Section~\ref{subsec:propositions} specify when role creation is the appropriate response and when concentrated or distributed extension remains more suitable.

\subsection{Organizational Placement of the CAIO}
For the CAIO to perform the role theorized in this paper effectively, the position cannot be structured as a technical specialist or subordinate innovation function \citep{Hambrick1984UpperManagers, Banker2011CIOPerformance1}. Its responsibilities span strategy, enterprise-wide coordination, and governance. Effective placement therefore requires proximity to the chief executive officer and inclusion in the top management team \citep{Krause2022TopManagementTeam}, where the role has the authority to shape firm-wide priorities and resolve conflicts across business units and functional domains.

This placement follows directly from the logic of the framework. AI affects competitive positioning, operating models, governance arrangements, and decision architectures at the same time. The CAIO therefore requires enterprise-wide visibility and influence rather than narrow functional control. Positioning the role within the top management team gives it the structural standing to translate AI-related opportunities and risks into coherent organizational action. This argument builds on strategic leadership perspectives, which hold that major organizational outcomes are shaped by the attention, authority, and integrative capacity of top executives \citep{Hambrick1984UpperManagers, Carpenter2004UpperEchelons, Neely2020Metacritiques}. Research likewise shows that the reporting structure of senior technology leaders shapes their strategic positioning and organizational influence \citep{Banker2011CIOPerformance1}. More recent research on technology-oriented executives suggests that their organizational effects depend on expertise, on their presence in the top management team, on reporting relationships, and on their ability to direct attention across the firm \citep{Bendig2022WhenPresence}.

\subsection{Propositions}
\label{subsec:propositions}

This subsection states the framework's central predictions as four propositions. Each is stated so that it could in principle be disconfirmed.

The mechanism developed in Section~\ref{subsec:mechanism} holds that AI generates a judgment-governance problem when three properties are jointly present: distributed accountability for judgment, upstream governance, and non-stationarity. The properties are not substitutes. Distributed accountability without upstream governance produces a compliance question that a legal or risk function can absorb. Upstream governance without non-stationarity produces a one-time architectural question that the CIO or CDO can absorb. Non-stationarity without distributed accountability produces an operational monitoring question that an engineering function can absorb. Only when the three coincide does the problem exceed the absorptive capacity of any single existing role and make the cost of a dedicated role worth bearing.

\begin{quote}
\textbf{Proposition 1 (Emergence).} \textit{Firms are more likely to establish a dedicated CAIO the more their use of AI exhibits distributed accountability for judgment, upstream governance demands, and non-stationarity. The three properties are complements: each strengthens the effect of the others.}
\end{quote}

The complementarity is the substantive claim. An additive prediction would follow from a view in which each property independently raises the demand for AI leadership. The framework predicts instead that the properties are complementary, because it is their joint presence that defeats absorption by an existing role. The proposition is therefore distinct from a general claim that more AI leads to more AI leadership. It is also testable, since the interaction term should carry explanatory weight beyond the main effects. This logic extends prior work showing that technology-oriented executive roles enter the top management team when technological change reshapes strategic priorities and cuts across functional boundaries \citep{Bendig2022WhenPresence, Gao2025ChiefInnovation}, while specifying the particular configuration of AI properties whose joint presence makes a dedicated role worthwhile.

A firm that elevates AI to the executive level still faces a choice among the three configurations developed in Section~\ref{subsec:two-responses}. The framework predicts that this choice is shaped less by the intensity of the AI problem than by the executive structure the problem encounters. Two features of that structure matter most. The first is the state of the incumbent CDO mandate. Where a CDO is present and actively expanding its remit, AI is the natural next addition to that remit, and concentrated extension is the path of least resistance. The second is the degree to which AI initiatives are coupled to existing data and digital infrastructure. Where AI builds directly on infrastructure an incumbent already controls, that incumbent can absorb AI without ceding authority. Where AI cuts across infrastructures held by several executives, no incumbent can absorb it alone.

\begin{quote}
\textbf{Proposition 2 (Form).} \textit{Among firms that elevate AI to the executive level, the form of AI leadership depends on the structure of incumbent authority and infrastructure coupling:}\\[2pt]
\textit{(a) concentrated extension is more likely where an incumbent executive holds an expanding adjacent mandate and AI is tightly coupled to that executive's infrastructure;}\\[2pt]
\textit{(b) distributed extension is more likely where authority over AI-relevant infrastructure is divided among several strong incumbents;}\\[2pt]
\textit{(c) role creation is more likely where no incumbent holds an expanding adjacent mandate or where AI initiatives are weakly coupled to any single incumbent's infrastructure.}
\end{quote}

The proposition predicts variation in form that does not track the severity of the underlying AI problem. Two firms facing identical judgment-governance problems may adopt different configurations because they inherit different executive structures. The framework thus explains why some firms create a CAIO, others extend the CDO, and others govern AI through federated structures. It does not treat any single configuration as the uniquely correct response to a given level of AI demand.

Creating a dedicated role does not by itself resolve the judgment-governance problem. The accountability externality developed in Section~\ref{subsec:design-problems} is resolved only when one actor internalizes the costs that functions would otherwise externalize, and internalization requires authority. A CAIO without authority over enterprise-wide AI decisions, without a budget that lets the role commit resources, and without a seat where firm-wide trade-offs are made cannot internalize the externality. Such a role can name the problem but cannot resolve it, and may worsen jurisdictional ambiguity by adding another partial claimant to contested territory.

\begin{quote}
\textbf{Proposition 3 (Effectiveness).} \textit{A dedicated CAIO improves the coherence of enterprise AI governance to the extent that the role holds top-management-team membership, budget authority, and formal decision rights over AI. Absent these conditions, the role is unlikely to improve governance coherence and may reduce it by adding another locus of partial and overlapping authority.}
\end{quote}

The proposition makes effectiveness conditional on the role's structural endowment, not on its existence or title. It separates symbolic appointments from consequential ones and identifies the conditions a firm must meet for a CAIO to deliver the coordination the framework attributes to the role. It also implies a failure mode. A CAIO created without authority is not merely ineffective but potentially counterproductive, because it formalizes a claim to authority it cannot exercise.

Extension configurations preserve the architecture of the roles they extend. Section~\ref{subsec:two-responses} identified three failure modes that follow from this preservation: governance lag, capability dilution, and legitimacy ambiguity. These failure modes accumulate as AI's organizational footprint grows, because the gap between the judgment-governance problem and the integration toolkit of an extended role widens with AI's centrality. Extension is therefore predicted to be transitional under sustained AI growth, not a stable response. Where AI's centrality recedes, the same logic runs in reverse, and the extended mandate is predicted to shed its AI component and revert toward its legacy form. The dedicated role, once created, faces an evolutionary question of its own: whether the conditions that justified its creation continue to hold, or whether organizational learning and maturing governance infrastructure gradually routinize the judgment-governance problem the role was built to manage.

\begin{quote}
\textbf{Proposition 4 (Evolution).}\\[2pt]
\textit{(a) Extension configurations are more likely to evolve toward dedicated role creation as AI becomes more organizationally central, and to revert toward legacy mandates as its centrality declines.}\\[2pt]
\textit{(b) Once created, a dedicated CAIO is more likely to persist while distributed accountability for judgment, upstream governance demands, and non-stationarity remain high, and more likely to be absorbed into incumbent mandates as governance infrastructure, organizational learning, and regulatory standardization routinize these demands.}
\end{quote}

Part (a) treats the hybrid titles observed in current practice, such as Chief Data and AI Officer or Chief Digital and AI Officer, as a transitional stage. Transitions toward role creation should be observable over time and should concentrate in regulated, data-intensive, and AI-native settings, where AI's distinctive properties are most strongly expressed; reversions should concentrate where AI recedes to the periphery of the business model. Part (b) extends the evolutionary logic to the dedicated role itself. Persistence is conditional: the properties that make the role worth creating are also what sustain it, and once the judgment-governance problem becomes routinized enough for incumbent mandates to absorb, the conditions that justify a dedicated role no longer hold. The framework thereby specifies the conditions under which accounts of the CAIO as a temporary transformation role hold and the conditions under which they fail.

\subsection{Scope Conditions}

The framework applies when AI becomes consequential enough to require an explicit allocation of executive authority. Where AI remains confined to discrete applications within a single function, the judgment-governance problem developed in Section~\ref{subsec:mechanism} does not arise at the executive level. This boundary is a scope condition that precedes the propositions.

Even within that scope, a new executive arrangement is not always necessary. AI may be strategically important without all three properties intensifying together. Its behavior may remain stable, or its use across functions may create limited accountability exposure. Where these properties do not jointly generate demands that exceed existing mandates, established roles remain sufficient, as Proposition~1 implies. The framework therefore explains both when firms need to reconsider their executive arrangements and when they can accommodate AI within existing ones.

The framework also does not assume that firms choose the most suitable configuration. Executive politics, imitation of peers, and the symbolic appeal of an AI title can produce arrangements poorly matched to organizational needs, with governance costs anticipated by Proposition~3. The framework's claims are strongest in regulated, data-intensive, and AI-native settings and weakest where AI remains peripheral to value creation.

\section{Empirical Evidence on Executive AI Leadership}
\label{sec:evidence}

The framework predicts systematic variation in how organizations allocate executive authority over AI. This section documents that the phenomenon the framework explains is real, accelerating, and heterogeneous, drawing on what is, to my knowledge, the first systematic archival census of executive AI leadership. Four linked analyses examine the diffusion of executive AI configurations, the organizational conditions under which firms elevate AI to the executive level and the form that elevation takes, the tenure and early evolution of the resulting roles, and the mandates organizations attach to them. The evidence documents organizational patterns and informs the propositions, but does not directly test their underlying mechanisms. The propositions specify mechanisms that title-based archival data cannot measure directly, such as the joint intensity of AI's three properties and the absorptive capacity of incumbent mandates. Section~\ref{subsec:correspondence} states these implications and their limits.

\subsection{Data and Measurement}
\label{subsec:data}

I assemble the census from three independent source systems. First, BoardEx organization composition files (North America, Europe, United Kingdom, Rest of World) provide role spells---title, start date, end date, and seniority---for board members and senior managers. A broad search for titles containing the standalone term ``AI'' or the phrase ``artificial intelligence'' returned 754 role spells, which a fixed classification rule sorts into the paper's configurations: a \textit{dedicated CAIO} (Chief AI Officer with no other domain in the title; role creation), a \textit{hybrid title} (an AI mandate fused with data, digital, technology, product, or analytics domains, such as Chief Data \& AI Officer; concentrated extension), and a senior sub-executive \textit{Head of AI}. Research titles and non-executive AI positions are excluded. The rule yields 424 executive AI spells across 374 organizations between 2010 and August 2026; the Appendix reports the full sample construction and classification rule. Second, S\&P Capital IQ Key Developments provides 141 dated appointment announcements covering 134 organizations, seven of which have two announcements. These texts provide the material for the mandate-content analysis. Third, full-text search of the U.S. Securities and Exchange Commission's EDGAR filing system counts regulatory filings mentioning the title. US analyses link BoardEx organizations to Compustat and CRSP through the Wharton Research Data Services (WRDS) link tables.

Three measurement boundaries follow from the design. Distributed (federated) extension produces no distinctive executive title, so organizations governed that way fall into the non-adopter category. The census therefore identifies role creation and concentrated extension but cannot separately identify distributed extension. Title-based classification cannot observe reporting lines or decision rights directly; BoardEx's seniority classification and Execucomp provide partial proxies. Finally, because most spells begin after 2023, duration statistics are heavily right-censored, and all survival estimates use Kaplan--Meier methods.

\subsection{The Rise of Executive AI Leadership}
\label{subsec:rise}

\begin{figure}[htbp]
\centering
\caption{Executive AI Leadership Across Three Independent Data Systems}
\label{fig:triangulation}
\includegraphics[width=0.8\textwidth]{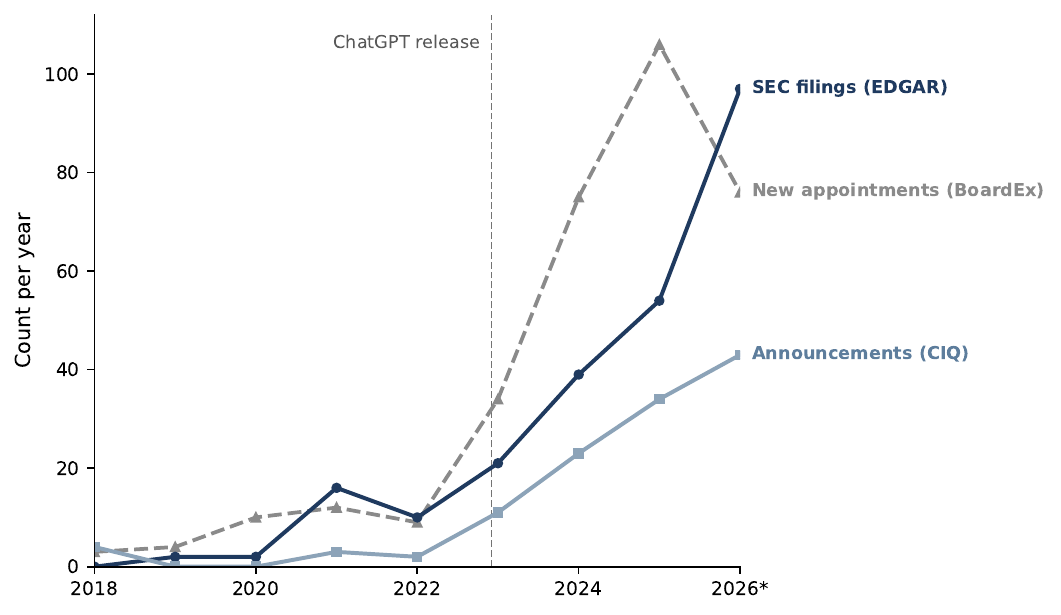}
\par\vspace{4pt}
{\footnotesize\raggedright\setlength{\parindent}{0pt}%
\textit{Note.} Annual counts of new executive AI appointments (BoardEx, dedicated $+$ hybrid), CAIO appointment announcements (S\&P Capital IQ Key Developments), and SEC filings mentioning the title (EDGAR full-text search). Dashed line marks the release of ChatGPT (November 2022). *2026 values cover January--July; the BoardEx down-tick reflects partial-year coverage, not a decline.\par}
\end{figure}

Figure~\ref{fig:rise} documented the diffusion: from single digits before 2020 to 29 organizations with a chief-level AI executive in 2022 and 258 by August 2026---127 with a dedicated CAIO and 134 with a hybrid title, three holding both---plus 58 with a senior Head of AI, 310 organizations in all. Two features stand out. The inflection is sharp and dates to 2023, coinciding with the diffusion of generative AI. And role creation and concentrated extension are expanding at almost the same pace, reflecting the variation the framework seeks to explain: organizations facing the same technological shock are institutionalizing executive AI authority through structurally different designs. Figure~\ref{fig:triangulation} triangulates the flow across three independent systems---new BoardEx appointments, Capital IQ appointment announcements, and SEC filings mentioning the title. All three series show a marked increase from 2023 onward; the pattern is not an artifact of any single database's coverage.

\begin{figure}[htbp]
\centering
\caption{Sector Concentration of Executive AI Leadership}
\label{fig:sectors}
\includegraphics[width=0.88\textwidth]{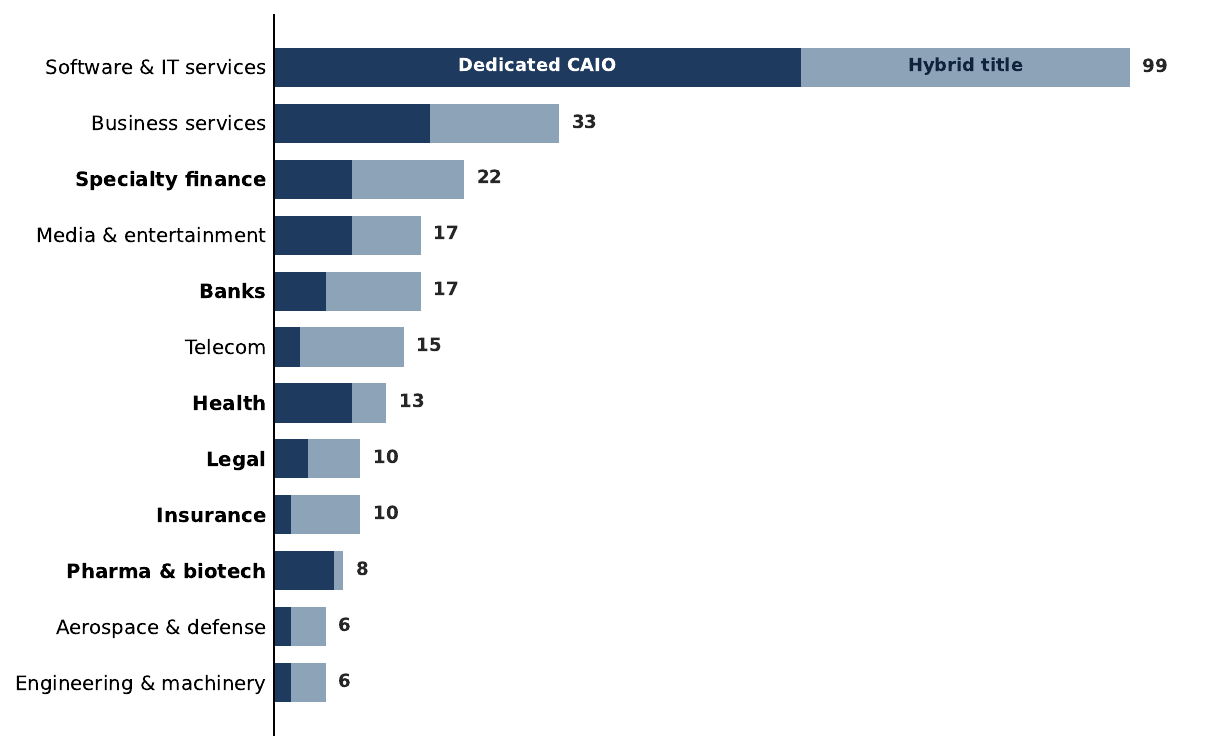}
\par\vspace{4pt}
{\footnotesize\raggedright\setlength{\parindent}{0pt}%
\textit{Note.} Organizations with executive AI leadership by BoardEx sector, 2010--2026, all regions; twelve largest sectors shown, bar-end values give organization counts. Each organization is classified by its first configuration, so segments partition organizations. Bold sector labels mark regulated or high-accountability sectors.\par}
\end{figure}

Figure~\ref{fig:sectors} shows where the role concentrates. Software and IT services lead, followed by business services; regulated, high-accountability sectors---specialty finance, banks, health, legal, insurance, pharmaceuticals---account for six of the twelve largest sectors, settings where the framework expects AI's governance demands to be most acute. Business-services organizations adopt dedicated CAIOs at notably high rates, consistent with judgment-intensive settings generating the strongest demand for executive AI accountability.

One further fact locates the role within the executive hierarchy: across the entire history of Execucomp, only nine named-executive-officer titles reference AI, and only 10\% of dedicated-CAIO spells in BoardEx carry board-level seniority. The role has proliferated as a senior title while remaining almost entirely absent from the proxy-statement top five. The role's structural endowment has not kept pace with its diffusion. Proposition~3 identifies this gap as consequential for effectiveness.

\subsection{Who Elevates AI, and in What Form}
\label{subsec:whoelevates}

Table~\ref{tab:desc} reports descriptive statistics for the adoption panel; Table~\ref{tab:models} examines adoption on a panel of 48,882 US firm-years (2016--2025, 70 first appointments), restricted to Compustat firms with a BoardEx link so that non-adoption is observed rather than imputed. R\&D intensity is R\&D expense over total assets, with missing R\&D expense set to zero and the ratio bounded between 0 and 1; no other outlier treatment is applied. Of the 101 Compustat-linked adopters, 74 adopt within the panel window, 70 of which have a panel observation in the adoption year and enter Panel A as events; 97 have fundamentals within two years of adoption and enter Panel B. Elevation of AI to the executive level is strongly predicted by firm size (odds ratio 1.60 per log-asset unit, $p<0.001$), technology-industry membership (OR 5.18, $p<0.001$), and, most notably, the presence of an incumbent Chief Data or Chief Digital Officer (OR 2.49, $p=0.011$). This association is consistent with the framework's argument that firms' responses to AI depend partly on the executive structures they inherit. It also points to continuity between successive waves of technology leadership: structures established for data and digital transformation may provide the foundation for AI leadership. After accounting for firm size and technology-industry membership, regulated-industry membership is not significantly associated with adoption. The sector concentration in Figure~\ref{fig:sectors} therefore reflects the composition of adopters rather than a separate effect of regulation.

Among firms that appoint an AI executive, however, the incumbent-role indicators do not explain the choice between dedicated and hybrid roles (Table~\ref{tab:models}, Panel B). Among 97 US-linked adopters, firms with an incumbent CDO chose the dedicated form slightly \textit{more} often (56\%) than firms without one (43\%), a pattern opposite to what a simple incumbency-based interpretation of Proposition~2 would suggest, though the difference is not statistically significant. Measurement is one interpretation: title-based CDO indicators capture presence, not the mandate expansion the proposition specifies. But the pattern also suggests a refinement worth theorizing: an incumbent CDO may absorb AI, producing a hybrid title, or may \textit{defend} an existing data mandate while a separate CAIO is created beside it. In the second case, role creation complements digital incumbency instead of replacing it. The observed pattern does not support a simple account in which CDO presence channels firms into extension. The choice between dedicated and hybrid roles may depend on how incumbent mandates evolve, which static title indicators do not capture.

\begin{table}[htbp]
\centering
\caption{Descriptive Statistics and Correlations}
\label{tab:desc}
\fontsize{9}{10.5}\selectfont\renewcommand{\arraystretch}{1.15}
\begin{tabularx}{\linewidth}{@{}>{\raggedright\arraybackslash}X cc cccccc@{}}
\toprule
Variable & $M$ & $SD$ & 1 & 2 & 3 & 4 & 5 & 6 \\ \midrule
\multicolumn{9}{@{}l}{\textit{Panel A: Adoption panel ($N=48{,}882$ US firm-years, 2016--2025)}} \\
1. First executive AI appointment ($\times 100$) & 0.14 & 3.78 & --- &  &  &  &  &  \\
2. Firm size (ln assets) & 6.58 & 2.65 & .04 & --- &  &  &  &  \\
3. R\&D intensity & 0.09 & 0.19 & $-$.01 & $-$.45 & --- &  &  &  \\
4. Regulated industry & 0.39 & 0.49 & $-$.01 & .05 & .21 & --- &  &  \\
5. Technology industry & 0.13 & 0.34 & .03 & $-$.06 & .05 & $-$.32 & --- &  \\
6. Incumbent CDO present & 0.03 & 0.17 & .03 & .20 & $-$.06 & $-$.01 & $-$.01 & --- \\[4pt]
\multicolumn{9}{@{}l}{\textit{Panel B: Form-choice sample ($N=97$ adopters, measured in the adoption year)}} \\
1. Dedicated form ($=$1) vs.\ hybrid ($=$0) & 0.45 & 0.50 & --- &  &  &  &  &  \\
2. Incumbent CDO at adoption & 0.19 & 0.39 & .10 & --- &  &  &  &  \\
3. Firm size (ln assets) & 8.99 & 2.66 & $-$.12 & .33 & --- &  &  &  \\
4. Regulated industry & 0.26 & 0.44 & $-$.06 & .20 & .24 & --- &  &  \\
5. Technology industry & 0.36 & 0.48 & .13 & $-$.08 & $-$.13 & $-$.44 & --- & \\ \bottomrule
\end{tabularx}
\par\vspace{4pt}\par\noindent
{\fontsize{8.5}{10}\selectfont\raggedright \textit{Notes.} $M$ $=$ mean; $SD$ $=$ standard deviation. Panel A: Compustat firms linked to BoardEx. First executive AI appointment is coded 100 in the year of a firm's first dedicated or hybrid appointment and 0 otherwise; the sample contains 70 such appointments. Regulated industry, technology industry, and incumbent CDO present are binary indicators. Incumbent CDO present identifies firms with a Chief Data Officer or Chief Digital Officer recorded in BoardEx in the observation year. Panel B: the US Compustat-linked adopters that enter Table~\ref{tab:models}, Panel B, with firm characteristics taken from the nearest fiscal year within two years before adoption; dedicated form equals 1 for a dedicated CAIO and 0 for a hybrid title; incumbent CDO at adoption equals 1 if a Chief Data Officer or Chief Digital Officer was in office on June 30 of the adoption year.\par}
\end{table}

\begin{table}[htbp]
\centering
\caption{Configuration Choice: Adoption and Form}
\label{tab:models}
\fontsize{9}{10.5}\selectfont\renewcommand{\arraystretch}{1.15}
\begin{tabularx}{\linewidth}{@{}>{\raggedright\arraybackslash}X cccc@{}}
\toprule
Variable & Coef. & SE & Odds ratio & $p$ \\ \midrule
\multicolumn{5}{@{}l}{\textit{Panel A: Discrete-time logit, first executive AI appointment (dedicated or hybrid)}} \\
Firm size (ln assets) & 0.47 & 0.06 & 1.60 & $<$0.001 \\
R\&D intensity & 0.73 & 1.26 & 2.07 & 0.565 \\
Regulated industry & $-$0.29 & 0.32 & 0.75 & 0.368 \\
Technology industry & 1.64 & 0.31 & 5.18 & $<$0.001 \\
Incumbent CDO present & 0.91 & 0.36 & 2.49 & 0.011 \\[4pt]
\multicolumn{5}{@{}l}{\textit{Panel B: Form choice at adoption, dedicated ($=$1) vs.\ hybrid ($=$0)}} \\
Incumbent CDO at adoption & 0.86 & 0.58 & 2.36 & 0.143 \\
Firm size (ln assets) & $-$0.12 & 0.09 & 0.89 & 0.166 \\
Regulated industry & $-$0.01 & 0.56 & 0.99 & 0.980 \\
Technology industry & 0.55 & 0.49 & 1.74 & 0.256 \\
Constant & 0.52 & 0.79 & --- & 0.509 \\ \bottomrule
\end{tabularx}
\par\vspace{4pt}\par\noindent
{\fontsize{8.5}{10}\selectfont\raggedright \textit{Notes.} Panel A examines first executive AI appointments using $N=48{,}882$ firm-year observations from 6,801 US Compustat firms linked to BoardEx during 2016--2025, so that non-adoption is observed rather than imputed. The sample includes 70 first appointments. Firms leave the sample after their first appointment. The model includes grouped year fixed effects, with standard errors clustered by firm; pseudo-$R^2=0.225$. Panel B compares dedicated and hybrid appointments among $N=97$ US Compustat-linked adopters, 44 of them dedicated. Dedicated roles account for 43\% of appointments among firms without an incumbent CDO and 56\% among firms with one; this difference is not statistically significant. Incumbent CDO refers to a Chief Data Officer or Chief Digital Officer recorded in BoardEx. Sources: BoardEx, Compustat, CRSP, and WRDS linking tables.\par}
\end{table}

\subsection{Tenure and Early Evolution}
\label{subsec:tenure}

\begin{figure}[htbp]
\centering
\caption{Tenure and Early Evolution of Executive AI Configurations}
\label{fig:evolution}
\includegraphics[width=\textwidth]{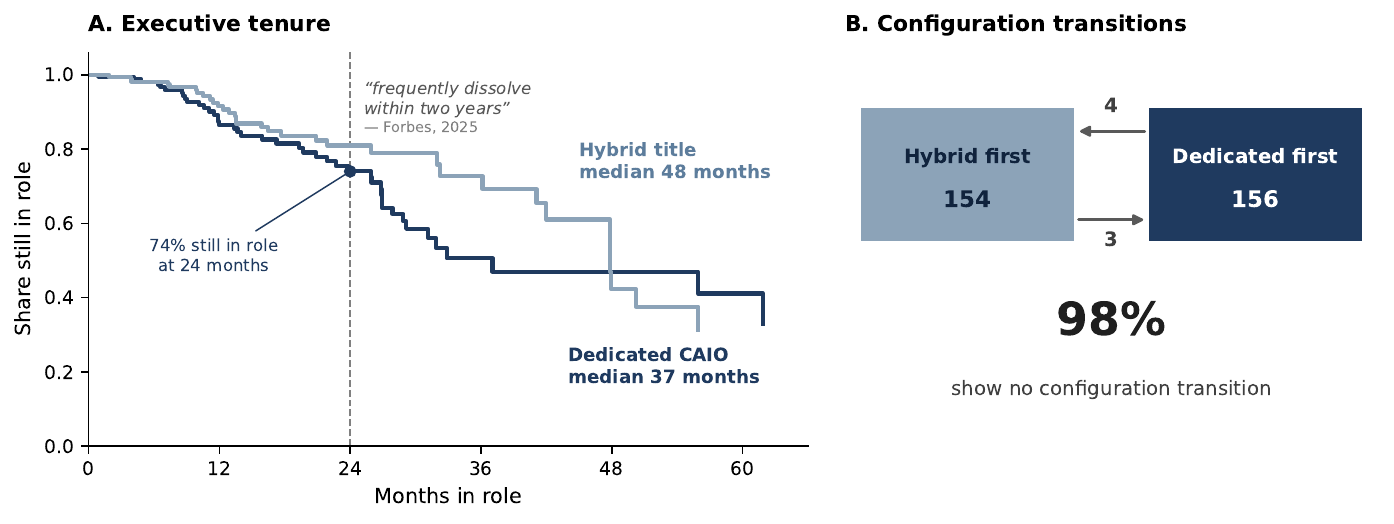}
\par\vspace{4pt}
{\footnotesize\raggedright\setlength{\parindent}{0pt}%
\textit{Note.} Panel A: Kaplan--Meier survival curves for 350 executive AI role spells (dedicated $n=177$, 43 completed; hybrid $n=173$, 33 completed; remainder right-censored at August 2026). The dashed line at 24 months marks the press claim that CAIO roles ``frequently dissolve within two years'' \citep{Marr2025AICrisis}; the marked point shows that 74\% of dedicated CAIOs remain in role at that horizon. Panel B: configuration transitions among the 310 organizations whose first chief-level AI role is a dedicated or hybrid title; a transition is a later title of the other type, and organizations whose chief-level role has lapsed without a recorded successor (52) count as non-transitions. The median observation window is roughly two years, so transitions are heavily right-censored.\par}
\end{figure}

Figure~\ref{fig:evolution} turns to dynamics. Kaplan--Meier estimates over the 350 chief-level spells (the 424 classified spells less 74 Head of AI spells) put median executive tenure at 37 months for dedicated CAIOs and 48 months for hybrid titles; roughly three-quarters of dedicated CAIOs remain in role at 24 months. These estimates contradict a press narrative in which CAIO positions ``frequently dissolve within two years'' \citep{Marr2025AICrisis}, a claim that, so far as I can establish, rests on no systematic evidence. The census supplies that evidence, and it points the other way. A position persists at least as long as its holder remains in it. The 74\% of dedicated CAIOs still in role at 24 months therefore implies that at most 26\% of dedicated positions could have been dissolved by that point. A na\"ive median computed on completed spells alone (16 months) reproduces the narrative almost exactly: that is the arithmetic of counting only the people who have already left, at a moment when three-quarters of all spells are still running. Dedicated-role holders do turn over faster than hybrid holders, but the difference is one of degree.

Configuration transitions remain rare: of the 310 organizations whose first executive AI role is a dedicated or hybrid title, 98\% show no transition between configurations to date (three organizations moved from a hybrid title to a dedicated CAIO, four the reverse). This measure records observed switches, not continuous occupancy: 52 of the 310 organizations (17\%) have no active chief-level AI role at the cutoff, which may reflect dissolution of the position, an unrecorded successor, or reporting lag, and title data cannot separate these. With a median observation window of roughly two years, it is too early to tell whether these configurations will remain stable. Assessing the transitions predicted by Proposition~4 requires longer follow-up. What the data establish is that neither mass dissolution nor mass conversion has occurred: as of 2026, role creation and concentrated extension are being institutionalized in parallel.

\subsection{What AI Executives Are Mandated to Do}
\label{subsec:functions}

\begin{figure}[htbp]
\centering
\caption{The Five Framework Functions in Appointment Announcement Language}
\label{fig:functions}
\includegraphics[width=0.85\textwidth]{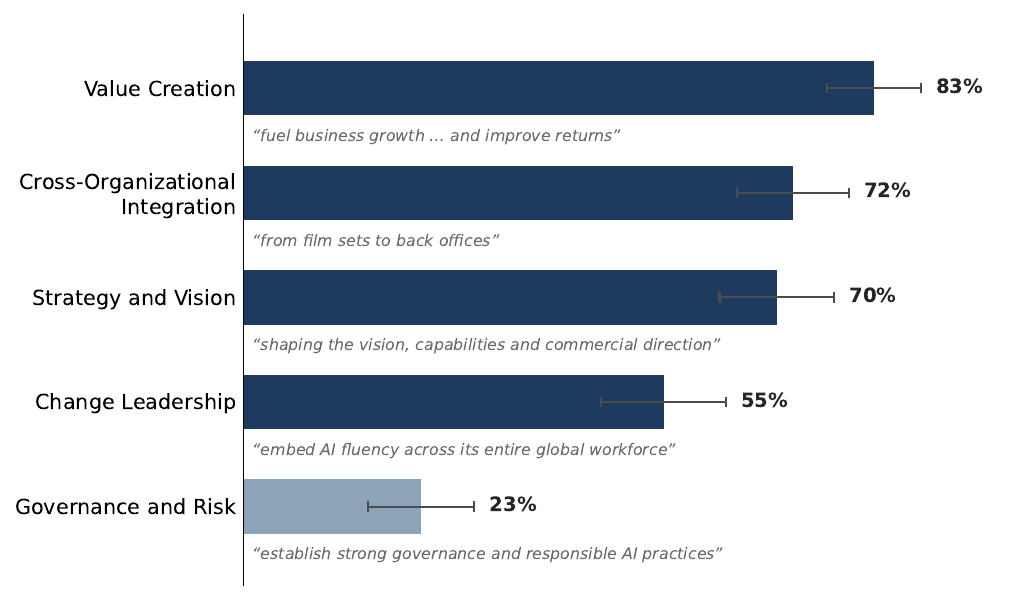}
\par\vspace{4pt}
{\footnotesize\raggedright\setlength{\parindent}{0pt}%
\textit{Note.} Share of all 141 CAIO/hybrid appointment announcements (S\&P Capital IQ Key Developments) whose text references each function, coded using the fixed keyword codebook reported in the Appendix; whiskers give 95\% confidence intervals. Italic fragments are verbatim excerpts from announcements in the sample. Governance \& risk (highlighted) is invoked in fewer than one in four announcements. Shares are similar for dedicated ($n=121$) and hybrid ($n=20$) announcements.\par}
\end{figure}

Figure~\ref{fig:functions} examines mandate content, coding all 141 appointment announcements (headline and situation text; 134 organizations) against the five functions the framework derives. The coding unit is the announcement, the denominator is 141, and the Appendix reports the keyword rules. Four of the five functions appear in a majority of announcements, so the functional architecture the framework proposes is visible in how organizations describe the role. The exception is the most theoretically consequential one. Value creation (83\%), cross-organizational integration (72\%), strategy and vision (70\%), and change leadership (55\%) dominate mandate rhetoric, while genuine governance-and-risk language---responsible AI, ethics, oversight, compliance---appears in fewer than one in four announcements (23\%). The function the framework identifies as the role's distinctive structural rationale---internalizing AI's accountability externalities---is the one organizations emphasize least when announcing the role. This gap between announced mandate and theorized design is consistent with a portion of appointments being oriented toward signaling and value narratives rather than the governance mandate that would make the role structurally consequential, and it gives the symbolic-versus-substantive distinction in Proposition~3 an observable form. The similarity of function shares across dedicated and hybrid announcements is itself informative: it is consistent with the framework's claim (Section~\ref{subsec:framework}) that the five functions are configuration-invariant demands, bundled differently across configurations rather than performed selectively.

\subsection{Implications, Robustness, and Limits}
\label{subsec:correspondence}

The findings inform the propositions without directly testing their underlying mechanisms. They also provide a baseline for examining how executive AI leadership evolves.

The main patterns hold across several robustness checks. Excluding appointments beginning in 2026 leaves 90 organizations with dedicated roles and 105 with hybrid roles at year-end 2025. Restricting the census to North America yields similarly balanced counts of 76 and 77. A stricter definition of dedicated titles excludes 19 organizations but preserves the broad pattern, with median tenures unchanged to one decimal place. Expanding the adoption model to all 79,750 Compustat firm-years, treating firms without a BoardEx link as non-adopters, raises the odds ratio for incumbent CDO presence to 4.00 and for technology-industry membership to 5.86. Table~\ref{tab:models} reports estimates for firms linked to BoardEx.

Several limitations qualify these findings. Titles cannot identify distributed extension or establish executives' actual mandates and authority. BoardEx coverage favors listed companies and large private organizations, limiting generalization to smaller firms. Recent appointments and ongoing tenures also restrict conclusions about long-term role persistence and configuration changes. Further research could address these limits through mandate documents, reporting-line data, and longer observation periods.

\section{Discussion}
 \label{sec:discussion}

\subsection{Theoretical Contributions}

This paper makes three contributions to research on executive leadership, organizational design, and AI governance.

First, it connects the design of executive roles to the properties of the technology being governed. Distributed accountability for judgment, upstream governance, and non-stationarity jointly create demands that can exceed existing executive mandates. This extends the explanation of technology leadership beyond the integration of systems and processes to the allocation of authority over judgments made by AI.

Second, the paper explains why firms facing similar technological pressures organize AI leadership differently. The choice among concentrated extension, distributed extension, and role creation depends on the executive authority and infrastructure firms inherit. A dedicated CAIO is thus one possible response to AI's governance demands, with existing arrangements remaining appropriate under identifiable conditions.

Third, the paper distinguishes the creation of a role from its effectiveness and persistence. The framework identifies five functions that all three configurations must perform, while the propositions explain why a dedicated CAIO requires authority, resources, and participation in firm-wide decisions to perform them effectively. They also explain when extended mandates may develop into dedicated roles and when dedicated roles may persist or be absorbed into existing mandates. Executive AI leadership is therefore treated as an organizational arrangement whose value and form can change over time.

\subsection{Implications for AI Governance and Executive Roles}

The framework carries three implications for research on AI governance and executive roles. First, it extends IT governance theory to a new governance object. That literature has long examined where authority over systems, data, and infrastructure should reside \citep{Sambamurthy1999Arrangements}. AI raises a distinct problem: the allocation of authority over machine judgment---what models are optimized for, which decisions they may make, and when human intervention is required. Judgment governance, as developed in Section~\ref{subsec:mechanism}, names these decisions about authority over AI judgments, which deployment-centered IT governance frameworks were not designed to address. The three configurations offer alternative ways to allocate that authority among executives. The announcement evidence highlights a related tension: governance and risk receive less emphasis than the other four functions when firms describe AI leadership appointments (Figure~\ref{fig:functions}), even though the framework identifies accountability as central to the role's purpose.

Second, the framework provides a general account of how executive technology roles emerge and take organizational form. Research on the CIO, CDO, and CISO has documented the antecedents and consequences of individual executive roles \citep{Bendig2022WhenPresence, Firk2021ChiefRole, Gao2025ChiefInnovation}. The framework's absorption-versus-differentiation logic explains when a new technology domain is incorporated into an incumbent mandate, distributed across existing executives, or differentiated into a new role, connecting role formation to the properties of the underlying technology and its organizational dependencies. This provides a common theoretical structure across otherwise role-specific literatures.

Third, the framework connects executive organization to the literature on algorithmic accountability and human oversight \citep{Salovaara2019HighOperations1, Grisold2025GuardrailsPredictability, Stelmaszak2025WhenUber}. Guardrails, delegation rules, and oversight practices specify how responsible AI operates at the level of individual systems and decisions. Executive role design determines the corresponding organizational structure: where accountability resides, who holds decision rights over AI, and who---if anyone---answers for judgments delegated to learned systems. The three configurations thus locate algorithmic accountability within the broader architecture of organizational authority.

\subsection{Managerial Implications}

For practitioners, the framework does not suggest that every organization needs a Chief AI Officer, but it identifies the conditions under which a dedicated executive role becomes warranted. The case for a CAIO is likely to be strongest where AI is strategically central, deployed across multiple functions, and associated with significant governance, regulatory, or reputational exposure. In such settings, dispersed ownership of AI may create fragmentation, duplicated efforts, and accountability gaps that are difficult to resolve within existing executive arrangements. The analysis also suggests that the effectiveness of the role will depend less on the title itself than on the clarity of its mandate, its relationship to adjacent C-suite roles, and its position within the broader executive structure. Organizations therefore need to determine whether AI leadership can be coordinated through existing roles or whether the scale and significance of AI warrant a more explicit concentration of executive authority. In this sense, the CAIO is one of several organizational responses to the growing strategic and governance demands of enterprise-wide AI.

\subsection{Research Agenda}

Table~\ref{tab:caio_research_agenda} organizes the research directions that follow.

\begin{table}[htbp]
\centering
\caption{A Research Agenda for Executive AI Leadership and the Chief AI Officer}
\label{tab:caio_research_agenda}
\fontsize{9}{10.5}\selectfont\renewcommand{\arraystretch}{1.15}
\begin{tabularx}{\linewidth}{@{}>{\raggedright}p{3.2cm}@{\quad}>{\raggedright\arraybackslash}p{5.4cm}@{\quad}>{\raggedright\arraybackslash}p{3.0cm}@{\quad}>{\raggedright\arraybackslash}X@{}}
\toprule
Research Domain & Research Questions & Theoretical Lenses & Methods \\ \midrule
Emergence and\newline configuration choice &
Under what conditions do firms create a dedicated CAIO, extend an incumbent mandate, or coordinate AI through committee structures?\newline How do AI strategic centrality, cross-functional deployment, and governance exposure jointly affect the configuration a firm adopts? &
Upper echelons theory\newline Contingency theory\newline TOE framework &
Archival panel studies; event-history models; matched-sample analyses; comparative case studies. \\[6pt]
Role design and\newline placement &
Which combinations of reporting line, budget, and decision rights enable a CAIO to integrate strategy, infrastructure, and governance?\newline When is a dedicated CAIO more effective than concentrated or distributed extension? &
Organizational design\newline Decision-rights theory\newline Role theory &
Multi-case studies; survey-based role-mapping; executive interviews; archival studies of reporting structures; social-network analysis. \\[6pt]
Jurisdiction and\newline role interactions &
How are mandates negotiated between the CAIO and adjacent executive roles?\newline When does overlap create complementarity versus conflict? &
Attention-based view\newline Coordination theory\newline Micropolitics &
Process tracing; executive interviews; meeting observations; multi-respondent surveys. \\[6pt]
Leadership\newline mechanisms &
What do CAIOs actually do to translate AI opportunities into enterprise action?\newline How do they prioritize use cases, govern experimentation, and scale AI across units? &
Socio-technical systems\newline Orchestration theory\newline Sensemaking &
Longitudinal cases; ethnography; archival and process data; project-level comparisons. \\[6pt]
Outcomes and\newline performance &
What organizational outcomes follow from each configuration?\newline Does configuration choice affect AI scaling speed, innovation quality, decision quality, risk incidents, or legitimacy? &
Dynamic capabilities\newline Digital Innovation\newline Legitimacy theory &
Difference-in-differences; event studies; innovation portfolio analysis; comparative case studies. \\[6pt]
Role evolution and\newline institutionalization &
Are extension configurations stable or transitional under sustained AI growth?\newline Under what conditions do firms transition between configurations, and how does the dedicated CAIO role itself evolve over time? &
Institutional theory\newline Path dependence\newline Role evolution &
Longitudinal archival analysis; repeated interviews; comparative institutional studies. \\[6pt]
Human--AI delegation and\newline decision rights &
How should authority be divided between executives and AI systems in strategic, operational, and compliance decisions?\newline When should human override or escalate? &
Human-AI interaction\newline Behavioral strategy\newline Delegation theory  &
Field studies of workflows; simulations; vignette experiments; behavioral experiments. \\[6pt]
Ecosystem and\newline sectoral variation &
How do executive AI configurations differ across public-sector, platform, multinational, and highly regulated contexts?\newline How do external platforms and partners reshape executive AI leadership? &
Institutional complexity\newline Institutional logics\newline Ecosystem theory &
Multi-level comparative cases; configurational analysis; embedded case research. \\ \bottomrule
\end{tabularx}
\par\vspace{4pt}\par\noindent
{\fontsize{8.5}{10}\selectfont\raggedright \textit{Notes.} CAIO $=$ Chief AI Officer; CIO $=$ Chief Information Officer; CTO $=$ Chief Technology Officer; CDO $=$ Chief Digital Officer.\par}
\end{table}

The framework opens three avenues for research. First, studies could examine how AI's strategic centrality, deployment across functions, and regulatory or reputational exposure shape firms' choice of executive configuration. Second, research could investigate how CAIO mandates are negotiated with adjacent executives, focusing on overlapping authority and the allocation of decision rights. Third, longitudinal studies could trace movement between configurations as AI's organizational importance changes, examining when extended mandates develop into dedicated roles and when dedicated roles persist, expand, or are absorbed into existing positions. Comparisons across industries and organizational settings would help establish the conditions under which these different trajectories occur.

\section{Conclusion}
 \label{sec:conclusion}

AI is prompting firms to reconsider how they allocate executive authority. The CAIO Framework explains why some create a dedicated role, while others extend existing mandates or distribute responsibility across executives. These choices depend on AI's governance demands and the executive structures firms inherit. Archival evidence documents the spread of executive AI roles and informs the theory through patterns of adoption, tenure, and announced responsibilities. A dedicated CAIO is therefore a conditional organizational response whose effectiveness depends on authority and whose persistence depends on continuing governance needs. The paper connects executive role design with empirical evidence on how firms organize AI leadership.

\singlespacing
\bibliographystyle{apalike}
\bibliography{bib_clean}

\clearpage

\appendix

\section{Empirical Sample Construction and Classification}
\label{app:sample}

Table~\ref{tab:sampleflow} reports the construction of the census. The classification rule is fixed and reproducible: titles containing ``chief,'' an AI token, and ``officer'' are classified as dedicated when no other domain token (data, digital, technology, information, analytics, strategy, product, and related terms) appears in the title, and as hybrid otherwise; ``Head of AI'' titles without a chief-officer form are classified as senior sub-executive positions; titles containing ``scientist'' and non-executive AI positions (vice presidents and directors of AI) are excluded. Table~\ref{tab:codebook} reports the five-function keyword codebook applied to announcement texts. Each announcement, comprising its headline and accompanying description, is treated as one observation (141 announcements covering 134 organizations). A function is recorded as present when at least one of its keyword patterns appears, and percentages are calculated across announcements. Matching runs over the full announcement text, including biographical and company passages, so the shares measure keyword prevalence and should be read as upper bounds on mandate content rather than as validated mandate codes.

\begin{table}[htbp]
\centering
\caption{Sample Construction and Coverage}
\label{tab:sampleflow}
\fontsize{9}{10.5}\selectfont\renewcommand{\arraystretch}{1.15}
\begin{tabularx}{\linewidth}{@{}>{\raggedright\arraybackslash}X r@{}}
\toprule
Item & Count \\ \midrule
BoardEx role spells matched by broad AI-title net (4 regions) & 754 \\
\quad less research titles & 4 \\
\quad less non-executive AI positions & 293 \\
\quad less missing or pre-2010 start dates & 33 \\
Classified executive AI spells, 2010--2026 & 424 \\
Organizations with at least one executive AI role & 374 \\
Organizations with active role, August 2026 (dedicated / hybrid / Head of AI; 310 unique) & 127 / 134 / 58 \\
S\&P Capital IQ Key Developments appointment announcements, classified (organizations covered) & 141 (134) \\
US Compustat-linked adopters (first dedicated or hybrid appointment) & 101 \\
Firm-year panel for the adoption model (2016--2025; Compustat firm-years with a BoardEx link) & 48{,}882 \\
Completed / right-censored executive spells & 76 / 274 \\
AI-titled named executive officers, Execucomp (all years) & 9 \\ \bottomrule
\end{tabularx}
\par\vspace{4pt}\par\noindent
{\fontsize{8.5}{10}\selectfont\raggedright \textit{Notes.} Sources: BoardEx, S\&P Capital IQ Key Developments, SEC EDGAR full-text search, Compustat, CRSP, Execucomp, via WRDS. All counts as of August 9, 2026, with BoardEx coverage through August 1, 2026.\par}
\end{table}

\begin{table}[htbp]
\centering
\caption{Five-Function Keyword Codebook for Appointment Announcements}
\label{tab:codebook}
\fontsize{9}{10.5}\selectfont\renewcommand{\arraystretch}{1.15}
\begin{tabularx}{\linewidth}{@{}>{\raggedright}p{3.4cm}@{\quad}>{\raggedright\arraybackslash}X@{}}
\toprule
Function & Keyword patterns \\ \midrule
Strategy and vision & strateg-, vision, roadmap, direction, foresight, long-term \\[3pt]
Value creation & value, growth, revenue, product, innovat-, monetiz-, commercial, customer, efficien- \\[3pt]
Cross-organizational integration & across, enterprise-wide, integrat-, coordinat-, scale/scaling, business unit(s), functions, organization-wide, group-wide \\[3pt]
Change leadership & transform-, change, culture, adopt-, upskill, training, talent, educat-, literacy \\[3pt]
Governance and risk & governance, ``responsible AI/use/deployment/development/innovation,'' responsibly, ethic-, ``AI risk,'' ``risk management,'' compliance, regulat-, trustworthy, ``AI safety,'' ``safe and/use/deployment,'' oversight, privacy \\ \bottomrule
\end{tabularx}
\par\vspace{4pt}\par\noindent
{\fontsize{8.5}{10}\selectfont\raggedright \textit{Notes.} The sample comprises $N=141$ appointment announcements covering 134 organizations from S\&P Capital IQ Key Developments. Each announcement includes its headline and accompanying description. A function is recorded as present when at least one of its keyword patterns appears anywhere in the announcement, regardless of capitalization. Function names correspond to those used in the text and Figure~\ref{fig:functions}. The word ``responsible'' alone does not count as evidence of governance and risk because it commonly appears in general descriptions of executive duties.\par}
\end{table}

\end{document}